\documentstyle[aps]{revtex}
\newcommand{\be}{\begin{equation}}
\newcommand{\ee}{\end{equation}}
\newcommand{\ba}{\begin{eqnarray}}
\newcommand{\ea}{\end{eqnarray}}
\newcommand{\baa}{\begin{eqnarray*}}
\newcommand{\eaa}{\end{eqnarray*}}

\newcommand{\dis}{\displaystyle}
\newcommand{\biq}{\mbox{\boldmath $q$}}
\newcommand{\bip}{\mbox{\boldmath $p$}}
\newcommand{\biL}{\mbox{\boldmath $\Lambda$}}
\newcommand{\bil}{\mbox{\boldmath $\lambda$}}

\begin{document}
\draft

{\pagestyle{empty}

{\renewcommand{\thefootnote}{\fnsymbol{footnote}}
\centerline{\large \bf Multidimensional replica-exchange method
for free-energy calculations}
}
\vskip 2.0cm
 
\centerline{Yuji Sugita,$^{a,}$\footnote{\ \ e-mail: sugita@ims.ac.jp}
Akio Kitao,$^{b,}$\footnote{\ \ e-mail: kitao@qchem.kuchem.kyoto-u.ac.jp}
and Yuko Okamoto$^{a,c,}$\footnote{\ \ e-mail: okamotoy@ims.ac.jp}}
\centerline{$^a${\it Department of 
Theoretical Studies, Institute for Molecular Science}}
\centerline{{\it Okazaki, Aichi 444-8585, Japan}}
\centerline{$^b${\it Department of 
Chemistry, Graduate School of Science, Kyoto University}}
\centerline{{\it Kyoto 606-8502, Japan}}
\centerline{$^c${\it Department of Functional Molecular Science,
The Graduate University for Advanced Studies}}
\centerline{{\it Okazaki, Aichi 444-8585, Japan}}

\vskip 2.0cm
\centerline{{\it J. Chem. Phys.} {\bf 113} (2000), in press.}
 
\begin{abstract}

We have developed a new simulation algorithm for free-energy
calculations.
The method
is a multidimensional extension of the 
replica-exchange method.  While pairs of replicas with different
temperatures are exchanged during the simulation in the original 
replica-exchange method, pairs of replicas with different temperatures
and/or different parameters of the potential energy
are exchanged in the new algorithm.
This greatly enhances the sampling of the conformational
space and allows accurate calculations of free energy in a wide
temperature range from a single simulation run, using the
weighted histogram analysis method.

\end{abstract}
}


\section{INTRODUCTION}
In complex systems such as a system of proteins, it is difficult
to obtain accurate canonical distributions at low temperatures by the
conventional molecular dynamics (MD) or Monte Carlo (MC)
simulations.
This is because there exist a huge number of local-minimum states
in the potential energy surface, and the simulations tend to get
trapped in one of the local-minimum states.
One popular way to overcome this difficulty is to perform a
{\it generalized-ensemble} simulation, which
is based on non-Boltzmann probability weight factors so that
a random walk in energy space may be realized
(for a review see \cite{RevHO}).
The random walk allows the simulation to go over any
energy barrier and sample much wider configurational space than
by conventional methods.
Monitoring the energy in a single simulation run, one can 
obtain not only
the global-minimum-energy state but also any thermodynamic
quantities as a function of temperature for a wide temperature
range.
The latter is made possible by the single-histogram \cite{FS1}
or multiple-histogram \cite{FS2} reweighting techniques (an
extension of the multiple-histogram method is also referred to
as the weighted histogram analysis method (WHAM) \cite{WHAM}).
  
Three of the most well-known generalized-ensemble methods are
perhaps {\it multicanonical algorithm} \cite{MUCA},
{\it simulated tempering} \cite{ST1,ST2}, and
{\it replica-exchange method} 
\cite{RE1}-\cite{RE5}.  (The replica-exchange
method is also referred to as
{\it replica Monte Carlo method} \cite{RE2},
{\it multiple Markov chain method} \cite{RE4}, and
{\it parallel tempering} \cite{RE5}.)
These algorithms have already been 
used in many applications in
protein and related systems (see, for instance,
Refs.~\cite{HO} - \cite{MO00} for multicanonical algorithm,
Refs.~\cite{IRB1} - \cite{HO96b} for simulated tempering, and
Refs.~\cite{H97} - \cite{Yama} for replica-exchange method).
    
The replica-exchange method (REM) has been drawing much 
attention recently because the probability
weight factors are essentially known {\it a priori}, whereas
they are not in most other generalized-ensemble algorithms
(and have to be determined by a tedius procedure).
In REM the generalized ensemble consists of noninteracting
copies (or replicas) of the original system
with different temperatures.  During a parallel MD or MC
simulation of each replica, a
pair of replicas are exchanged every few steps.
This procedure enforces random walks in the
replica (temperature) space.

In a previous work \cite{SO} we
worked out the details for the replica-exchange
molecular dynamics algorithm.
In this article we present a multidimensional
extension of the replica-exchange
method (similar generalizations of REM can also be
found in Refs. \cite{Huk2,YP}).
In the new algorithm, pairs of replicas with different temperatures
and/or different parameters of the potential energy
are exchanged.
As an example of the applications of the multidimensional
replica-exchange method, we discuss free-energy calculations
in detail.

The umbrella sampling method \cite{US} and free-energy perturbation
method, which is a special case of umbrella sampling,
have been widely used to calculate the
free energies in chemical processes \cite{US} - \cite{SK2}. 
In the umbrella sampling
method, a reaction coordinate is chosen and
free-energy profiles along the reaction coordinate are
calculated.
A series of independent simulations are performed to sample the
relevant range of the coordinate. To cover the entire range of the
coordinate, biasing potentials, which are called 
``umbrella potentials,'' are
imposed. Thus, the system is restrained to remain near the prechosen
value of the reaction coordinate specified by each umbrella potential,
and a series of simulations with different umbrella potentials are 
performed. 
WHAM \cite{WHAM} is often 
employed to calculate the free-energy
profiles from the histograms obtained by each simulation. 

Although the effectiveness of the umbrella sampling method is
well known, its successful implementation requires a careful
fine tuning.  For instance, the choice of 
the umbrella potentials is very important.
If the potentials are too strong, the conformational space
sampled by each simulation becomes quite narrow.
If the potentials are too weak, on the other hand, the system
does not remain near the prechosen value of the reaction coordinate.
The values of the coupling parameters $\lambda$
for the umbrella potentials should also be carefully chosen.
Various generalizations of the umbrella sampling method have thus
been introduced to sample the potential energy surface more 
effectively.  The $\lambda$-$dynamics$ \cite{Ldyn1} - \cite{Ike} is such
an example, where the coupling parameter $\lambda$ is treated
as a dynamical variable.  Another example is the 
{\it multicanonical
WHAM} \cite{ONHN}, which combines the umbrella sampling with
multicanonical algorithm.
In the present article we develop yet another generalization
of the umbrella sampling method (we refer to this method
as {\it replica-exchange umbrella sampling}), which is based on the
multidimensional extension of the replica-exchange method.

In Section II the multidimensional extension of the
replica-exchange method
is described in detail.  In particular, the replica-exchange
umbrella sampling method is introduced.
In Section III the results of the application of replica-exchange
umbrella sampling to a blocked alanine trimer are given.
Section IV is devoted to conclusions.

\section{METHODS}
Before we describe the {\it multidimensional replica-exchange method}
(MREM), let us briefly review the original {\it replica-exchange method}
(REM) \cite{RE1}-\cite{RE5} (see Ref.~\cite{SO} for details). 

We consider a system of $N$ atoms 
with their coordinate vectors and
momentum vectors denoted by 
$q \equiv \{{\biq}_1, \cdots, {\biq}_N\}$ and 
$p \equiv \{{\bip}_1, \cdots, {\bip}_N\}$,
respectively.
The Hamiltonian $H(q,p)$ of the system is the sum of the
kinetic energy $K(p)$ and the potential energy $E(q)$:
\begin{equation}
H(q,p) =  K(p) + E(q)~.
\label{eqn1}
\end{equation}
In the canonical ensemble at temperature $T$ 
each state $x \equiv (q,p)$ with the Hamiltonian $H(q,p)$
is weighted by the Boltzmann factor:
\begin{equation}
W_B(x) = e^{-\beta H(q,p)}~,
\label{eqn3}
\end{equation}
where the inverse temperature $\beta$ is defined by 
$\beta = 1/k_B T$ ($k_B$ is Boltzmann's constant). 

The generalized ensemble for REM consists of 
$M$ {\it noninteracting} copies (or, replicas) 
of the original system in the canonical ensemble
at $M$ different temperatures $T_m$ ($m=1, \cdots, M$).
We arrange the replicas so that there is always
exactly one replica at each temperature.
Then there is  a one-to-one correspondence between replicas
and temperatures; the label $i$ ($i=1, \cdots, M$) for replicas 
is a permutation of 
the label $m$ ($m=1, \cdots, M$) for temperatures,
and vice versa:
\begin{equation}
\left\{
\begin{array}{rl}
i &=~ i(m) ~\equiv~ f(m)~, \cr
m &=~ m(i) ~\equiv~ f^{-1}(i)~,
\end{array}
\right.
\label{eqn4b}
\end{equation}
where $f(m)$ is a permutation function of $m$ and
$f^{-1}(i)$ is its inverse.

Let $X = \left\{x_1^{[i(1)]}, \cdots, x_M^{[i(M)]}\right\} 
= \left\{x_{m(1)}^{[1]}, \cdots, x_{m(M)}^{[M]}\right\}$ 
stand for a ``state'' in this generalized ensemble.
Here, the superscript and the subscript in $x_m^{[i]}$
label the replica and the temperature, respectively.
The state $X$ is specified by the $M$ sets of 
coordinates $q^{[i]}$ and momenta $p^{[i]}$
of $N$ atoms in replica $i$ at temperature $T_m$:
\begin{equation}
x_m^{[i]} \equiv \left(q^{[i]},p^{[i]}\right)_m~.
\label{eqn5}
\end{equation}
Because the replicas are noninteracting, the weight factor for
the state $X$ in
this generalized ensemble is given by
the product of Boltzmann factors for each replica (or at each
temperature):
\begin{equation}
W_{REM}(X) = \exp \left\{- \dis{\sum_{i=1}^M \beta_{m(i)} 
H\left(q^{[i]},p^{[i]}\right) } \right\}
 = \exp \left\{- \dis{\sum_{m=1}^M \beta_m 
H\left(q^{[i(m)]},p^{[i(m)]}\right) }
 \right\}~,
\label{eqn7}
\end{equation}
where $i(m)$ and $m(i)$ are the permutation functions in 
Eq.~(\ref{eqn4b}).

We now consider exchanging a pair of replicas in the generalized
ensemble.  Suppose we exchange replicas $i$ and $j$ which are
at temperatures $T_m$ and $T_n$, respectively:  
\begin{equation}
X = \left\{\cdots, x_m^{[i]}, \cdots, x_n^{[j]}, \cdots \right\} 
\longrightarrow \ 
X^{\prime} = \left\{\cdots, x_m^{[j] \prime}, \cdots, x_n^{[i] \prime}, 
\cdots \right\}~. 
\label{eqn8}
\end{equation}
The exchange of replicas can be written in more detail as
\begin{equation}
\left\{
\begin{array}{rl}
x_m^{[i]} \equiv \left(q^{[i]},p^{[i]}\right)_m & \longrightarrow \ 
x_m^{[j] \prime} \equiv \left(q^{[j]},p^{[j] \prime}\right)_m~, \cr
x_n^{[j]} \equiv \left(q^{[j]},p^{[j]}\right)_n & \longrightarrow \ 
x_n^{[i] \prime} \equiv \left(q^{[i]},p^{[i] \prime}\right)_n~,
\end{array}
\right.
\label{eqn9}
\end{equation}
where the momenta are uniformly rescaled according to \cite{SO}
\begin{equation}
\left\{
\begin{array}{rl}
p^{[i] \prime} & \equiv \dis{\sqrt{\frac{T_n}{T_m}}} ~p^{[i]}~, \cr
p^{[j] \prime} & \equiv \dis{\sqrt{\frac{T_m}{T_n}}} ~p^{[j]}~.
\end{array}
\right.
\label{eqn11}
\end{equation}

In order for this exchange process to converge toward the equilibrium
distribution based on Eq.~(\ref{eqn7}), it is sufficient to 
impose the detailed balance
condition on the transition probability $w(X \rightarrow X^{\prime})$:
\begin{equation}
W_{REM}(X) \  w(X \rightarrow X^{\prime})
= W_{REM}(X^{\prime}) \  w(X^{\prime} \rightarrow X)~.
\label{eqn12}
\end{equation}
From Eqs.~(\ref{eqn1}), (\ref{eqn7}), (\ref{eqn11}), 
and (\ref{eqn12}), we have
\begin{equation}
\dis{\frac{w(X \rightarrow X^{\prime})} 
     {w(X^{\prime} \rightarrow X)}} 
= \exp \left( - \Delta \right)~,
\label{eqn13}
\end{equation}
where
\begin{eqnarray}
\Delta &=& \beta_m 
\left(E\left(q^{[j]}\right) - E\left(q^{[i]}\right)\right) 
- \beta_n
\left(E\left(q^{[j]}\right) - E\left(q^{[i]}\right)\right)~,
\label{eqn14a} \\
  &=& \left(\beta_m - \beta_n \right)
\left(E\left(q^{[j]}\right) - E\left(q^{[i]}\right)\right)~. 
\label{eqn14b}
\end{eqnarray}
This can be satisfied, for instance, by the usual Metropolis criterion
\cite{Metro}:
\begin{equation}
w(X \rightarrow X^{\prime}) \equiv
w\left( x_m^{[i]} ~\left|~ x_n^{[j]} \right. \right) 
= \left\{
\begin{array}{ll}
 1~, & {\rm for} \ \Delta \le 0~, \cr
 \exp \left( - \Delta \right)~, & {\rm for} \ \Delta > 0~.
\end{array}
\right.
\label{eqn15}
\end{equation}
Note that because of the velocity rescaling of Eq.~(\ref{eqn11})
the kinetic energy terms are cancelled out in Eqs.~(\ref{eqn14a})
(and (\ref{eqn14b})) and that 
the same criterion, Eqs.~(\ref{eqn14b}) and (\ref{eqn15}),
which was originally
derived for the Monte Carlo algorithm \cite{RE1}-\cite{RE5}
is recovered \cite{SO}.

A simulation of the 
{\it replica-exchange method} (REM) \cite{RE1}-\cite{RE5}
is then realized by alternately performing the following two
steps:
\begin{enumerate}
\item Each replica in canonical ensemble of the fixed temperature 
is simulated $simultaneously$ and $independently$
for a certain MC or MD steps. 
\item A pair of replicas,
say $x_m^{[i]}$ and $x_{n}^{[j]}$, are exchanged
with the probability
$w\left( x_m^{[i]} ~\left|~ x_{n}^{[j]} \right. \right)$ 
in Eq.~(\ref{eqn15}).
\end{enumerate}
In the present work, we employ the molecular dynamics algorithm
for step 1.
Note that in step 2 we exchange only pairs of replicas corresponding to
neighboring temperatures, because
the acceptance ratio of the exchange decreases exponentially
with the difference of the two $\beta$'s (see Eqs.~(\ref{eqn14b})
and (\ref{eqn15})).
Note also that whenever a replica exchange is accepted
in step 2, the permutation functions in Eq.~(\ref{eqn4b})
are updated.

The major advantage of REM over other generalized-ensemble
methods such as multicanonical algorithm \cite{MUCA}
and simulated tempering \cite{ST1,ST2}
lies in the fact that the weight factor 
is {\it a priori} known (see Eq.~(\ref{eqn7})), while
in the latter algorithms the determination of the
weight factors can be very tedius and time-consuming.
A random walk in ``temperature space'' is
realized for each replica, which in turn induces a random
walk in potential energy space.  This alleviates the problem
of getting trapped in states of energy local minima.

We now present our multidimensional extension of REM, which
we refer to as the {\it multidimensional replica-exchange method}
(MREM).
The crucial observation that led to the new algorithm is:  
As long as we have $M$ {\it noninteracting}
replicas of the original system, the Hamiltonian 
$H(q,p)$ of the system does not have to be identical
among the replicas and it can depend on a parameter
with different parameter values for different replicas.
Namely, we can write the Hamiltonian for the $i$-th
replica at temperature $T_m$ as
\begin{equation}
H_m (q^{[i]},p^{[i]}) =  K(p^{[i]}) + E_{\lambda_m} (q^{[i]})~,
\label{Eqn16}
\end{equation}
where the potential energy $E_{\lambda_m}$ depends on a
parameter $\lambda_m$ and can be written as
\begin{equation}
E_{\lambda_m} (q^{[i]}) = E_0 (q^{[i]}) + \lambda_m V(q^{[i]})~.
\label{Eqn16p}
\end{equation}
This expression for the potential energy is often used in
simulations.
For instance, in umbrella sampling \cite{US}, $E_0(q)$ and
$V(q)$ can be respectively taken as the original potential
energy and the ``biasing'' potential energy with the
coupling parameter $\lambda_m$.  In simulations of spin 
systems, on the other hand, 
$E_0(q)$ and $V(q)$ (here, $q$ stands for spins)
can be respectively considered as the
zero-field term and the magnetization term coupled with
the external field $\lambda_m$. 

While replica $i$ and temperature
$T_m$ are in one-to-one correspondence
in the original REM,
replica $i$ and ``parameter set''
$\Lambda_m \equiv (T_m,\lambda_m)$ are in one-to-one
correspondence in the new algorithm.
Hence, the present algorithm can be considered as a
multidimensional extension of the original replica-exchange
method where the ``parameter space'' is one dimensional 
(i.e., $\Lambda_m = T_m$).
Because the replicas are noninteracting, the weight factor 
for the state $X$ in this new
generalized ensemble is again given by the product
of Boltzmann factors for each replica (see Eq.~(\ref{eqn7})):
\begin{equation}
W_{MREM}(X) = \exp \left\{- \dis{\sum_{i=1}^M \beta_{m(i)} 
H_{m(i)}\left(q^{[i]},p^{[i]}\right) } \right\}
 = \exp \left\{- \dis{\sum_{m=1}^M \beta_m 
H_m\left(q^{[i(m)]},p^{[i(m)]}\right) }
 \right\}~,
\label{Eqn19}
\end{equation}
where $i(m)$ and $m(i)$ are the permutation functions in 
Eq.~(\ref{eqn4b}).
Then the same derivation
that led to the original replica-exchange
criterion (Eq.~(\ref{eqn15})) follows, and we have the following
transition probability of replica exchange (see Eq.~(\ref{eqn14a})):
\begin{equation}
w(X \rightarrow X^{\prime}) \equiv
w\left( x_m^{[i]} ~\left|~ x_n^{[j]} \right. \right) 
= \left\{
\begin{array}{ll}
 1~, & {\rm for} \ \Delta \le 0~, \cr
 \exp \left( - \Delta \right)~, & {\rm for} \ \Delta > 0~,
\end{array}
\right.
\label{Eqn20}
\end{equation}
where
\begin{equation}
\Delta = \beta_m 
\left(E_{\lambda_m}\left(q^{[j]}\right) - 
E_{\lambda_m}\left(q^{[i]}\right)\right) 
- \beta_n
\left(E_{\lambda_n}\left(q^{[j]}\right) - 
E_{\lambda_n}\left(q^{[i]}\right)\right)~.
\label{Eqn21}
\end{equation}
Here, $E_{\lambda_m}$ and $E_{\lambda_n}$ are the
total potential energies (see Eq.~(\ref{Eqn16p})).
Note that we need to newly evaluate the potential
energy for exchanged coordinates,
$E_{\lambda_m} (q^{[j]})$ and $E_{\lambda_n} (q^{[i]})$,
because $E_{\lambda_m}$ and $E_{\lambda_n}$ are in general
different functions.
The method is particularly suitable for parallel
computers.  Because one can minimize the amount of information
exchanged among nodes, it is best to assign each replica to
each node (exchanging $T_m,~E_{\lambda_m}$ and $T_n,~E_{\lambda_n}$ 
among nodes
is much faster than exchanging coordinates and momenta).
This means that we keep track of the permutation function
$m(i;t)=f^{-1}(i;t)$ in Eq.~(\ref{eqn4b}) as a function
of MD step $t$ throughout the simulation.

For obtaining the canonical distributions,
the weighted histogram analysis method (WHAM) \cite{WHAM}
is particularly suitable.
Suppose we have made a single run of the present
replica-exchange simulation with $M$ replicas that correspond
to $M$ different parameter sets
$\Lambda_m \equiv (T_m,\lambda_m)$ ($m=1, \cdots, M$).
Let $N_m(E_0,V)$ and $n_m$
be, respectively, 
the potential-energy histogram and the total number of
samples obtained for the $m$-th parameter set
$\Lambda_m$.
The expectation value 
of a physical quantity $A$ for any
potential-energy parameter value $\lambda$ at
any temperature $T=1/k_B \beta$
is then given by \cite{FS2,WHAM}
\begin{equation}
<A>_{T,\lambda} \ = \frac{\dis{\sum_{E_0,V}
A(E_0,V) P_{T,\lambda} (E_0,V)}}
{\dis{\sum_{E_0,V} P_{T,\lambda} (E_0,V)}}~,
\label{eqn18}
\end{equation}
where 
\begin{equation}
P_{T,\lambda} (E_0,V)
= \left[\frac{\dis{\sum_{m=1}^M g_m^{-1} N_m(E_0,V)}} 
{\dis{\sum_{m=1}^M n_{m} g_m^{-1} e^{f_m-\beta_m E_{\lambda_m}}}} \right]
e^{-\beta E_{\lambda}}~,
\label{eqn19}
\end{equation}
and
\begin{equation}
e^{-f_m} = \sum_{E_0,V} P_{T_m,\lambda_m} (E_0,V)~.
\label{eqn20}
\end{equation}
Here, 
$g_m = 1 + 2 \tau_m$, 
and $\tau_m$ is the integrated
autocorrelation time at temperature $T_m$ with
the parameter value $\lambda_m$.
Note that the unnormalized probability distribution
$P_{T,\lambda}(E_0,V)$ and the ``dimensionless''
Helmholtz free energy $f_m$ in Eqs.~(\ref{eqn19}) and
(\ref{eqn20}) are solved self-consistently
by iteration \cite{FS2,WHAM}.  

We can use this new replica-exchange method for 
free-energy calculations.  We first describe the free-energy
perturbation case.  The potential energy is given by
\begin{equation}
E_{\lambda} (q) = E_I (q) + \lambda \left(E_F (q) - E_I (q)\right)~,
\label{Eqn22}
\end{equation}
where $E_I$ and $E_F$ are the potential energy for
a ``wild-type'' molecule and a ``mutated''
molecule, respectively.  Note that this equation has the same
form as Eq.~(\ref{Eqn16p}).

Our replica-exchange simulation is performed for $M$ replicas
with $M$ different values of the parameters
$\Lambda_m = (T_m,\lambda_m)$.
Since $E_{\lambda = 0} (q) = E_I (q)$ and
$E_{\lambda = 1} (q) = E_F (q)$, we should choose enough
$\lambda_m$ values distributed in the range between 0 and 1
so that we may have sufficient replica exchanges.
From the simulation, $M$ histograms $N_m (E_I,E_F-E_I)$, or
equivalently $N_m(E_I,E_F)$, are obtained.  The Helmholtz
free-energy difference of ``mutation'' at 
temperature $T$,
$\Delta F \equiv F_{\lambda = 1} - F_{\lambda = 0}$, can then
be calculated from 
\begin{equation}
\exp (-\beta \Delta F) = \frac{Z_{T,\lambda=1}} 
{Z_{T,\lambda=0}} =  
\frac{\dis{\sum_{E_I,E_F}
P_{T,\lambda=1} (E_I,E_F)}}
{\dis{\sum_{E_I,E_F}
P_{T,\lambda=0} (E_I,E_F)}} ~,
\label{Eqn24}
\end{equation}
where $P_{T,\lambda} (E_I,E_F)$ are obtained from the WHAM
equations of Eqs.~(\ref{eqn19}) and (\ref{eqn20}).

We now describe another free-energy calculation based on
MREM applied to umbrella sampling \cite{US},
which we refer to as 
{\it replica-exchange umbrella sampling} (REUS).
The potential energy is a generalization of Eq.~(\ref{Eqn16p})
and is given by
\begin{equation}
E_{\bil} (q) = E_0 (q) + \sum_{\ell = 1}^L
\lambda^{(\ell)} V_{\ell} (q)~,
\label{Eqn25}
\end{equation}
where $E_0(q)$ is the original unbiased potential, 
$V_{\ell}(q)$ ($\ell =1, \cdots, L$) are the 
biasing (umbrella) potentials, and $\lambda^{(\ell)}$ are the
corresponding coupling constants
($\bil = (\lambda^{(1)}, \cdots, \lambda^{(L)})$).
Introducing a ``reaction coordinate'' $\xi$,
the umbrella potentials are usually written as harmonic
restraints:
\begin{equation}
V_{\ell} (q) = k_{\ell} \left[ \xi (q) - d_{\ell} \right]^2~,
~(\ell =1, \cdots, L)~,
\label{Eqn26}
\end{equation}
where $d_{\ell}$ are the midpoints and $k_{\ell}$ are the
strengths of the restraining potentials.
We prepare $M$ replicas with $M$
different values of the parameters
$\biL_m = (T_m,\bil_m)$, and the replica-exchange
simulation is performed.  Since the umbrella potentials
$V_{\ell} (q)$ in Eq.~(\ref{Eqn26})
are all functions of the reaction coordinate
$\xi$ only, we can take the histogram
$N_m (E_0,\xi)$ instead of
$N_m (E_0,V_1, \cdots, V_L)$.
The WHAM equations of
Eqs.~(\ref{eqn19}) and (\ref{eqn20}) can then be written as
\begin{equation}
P_{T,\bil} (E_0,\xi)
= \left[\frac{\dis{\sum_{m=1}^M g_m^{-1} N_m(E_0,\xi)}} 
{\dis{\sum_{m=1}^M n_{m} g_m^{-1} e^{f_m-\beta_m E_{\bil_m}}}} \right]
e^{-\beta E_{\bil}}~,
\label{Eqn27}
\end{equation}
and
\begin{equation}
e^{-f_m} = \sum_{E_0,\xi} P_{T_m,\bil_m} (E_0,\xi)~.
\label{Eqn28}
\end{equation}
The expectation value of a physical quantity $A$ is now
given by (see Eq.~(\ref{eqn18}))
\begin{equation}
<A>_{T,\bil} \ = \frac{\dis{\sum_{E_0,\xi}
A(E_0,\xi) P_{T,\bil} (E_0,\xi)}}
{\dis{\sum_{E_0,\xi} P_{T,\bil} (E_0,\xi)}}~.
\label{Eqn29}
\end{equation}

The potential of mean force (PMF), or free energy as a function of
the reaction coordinate, of the original, unbiased system 
at temperature $T$ is given by
\begin{equation}
{\cal W}_{T,\bil = \{0\}} (\xi) = - k_B T \ln
\left[ \sum_{E_0} P_{T,\bil = \{0\}} (E_0,\xi) \right]~,
\label{Eqn30}
\end{equation}
where $\{0\} = (0, \cdots, 0)$.
In the examples presented below, replicas were chosen so that
the potential energy for each replica includes exactly one (or zero)
biasing potential. 

\section{RESULTS AND DISCUSSION}
One of the applications of MREM,
{\it replica-exchange umbrella sampling} (REUS), was tested
for the system of a blocked peptide, alanine trimer.
The N and C termini of the peptide were blocked with 
acetyl and N-methyl groups, respectively. Since the thermodynamic
behavior of this peptide was extensively studied by the conventional
umbrella sampling \cite{Boc1}, 
it is a good test case to examine the effectiveness of the new method.
All calculations were based on MD simulations, and the force field 
parameters were taken from the
all-atom version of AMBER \cite{AMBER} with a distance-dependent dielectric, 
$\epsilon=r$, which mimics the presence of solvent.  
The computer code developed in Refs.~\cite{SK,KHG}, which is based
on Version 2 of PRESTO \cite{PRESTO}, was used.
The temperature during the MD simulations was controlled by the
constraint method \cite{HLM,EM}.
The unit time step was set to 0.5 fs, and we made an MD
simulation of $4 \times 10^6$ time steps (or 2.0 ns)
for each replica, starting from an extended conformation. 
The data were stored every 20 steps (or 10 fs) for a total of 
$2 \times 10^5$ snapshots. 
(Before taking the data, we made regular canonical MD simulations
of 100 ps for thermalization.
For replica-exchange simulations an additional REM simulation of
100 ps was made for further thermalization.)

In Table I we summarize the parameters characterizing the replicas
for the simulations performed in the present
work. They are one original replica-exchange simulation (REM1),
two replica-exchange umbrella sampling simulations (REUS1 and REUS2),
and two conventional umbrella sampling simulations (US1 and US2). 
The purpose of the present simulations is to test the effectiveness
of the replica-exchange umbrella sampling with respect to the conventional
umbrella sampling (REUS1 and REUS2 versus US1 and US2).
The original replica-exchange simulation without umbrella potentials
(REM1) was also made to set a reference standard for comparison.
For REM1,
replica exchange was tried every 20 time steps (or 10 fs),
as in our previous work \cite{SO}.  
For REUS simulations, on the other hand,
replica exchange was tried every 400 steps (or 200 fs),
which is less frequent than in REM1.
This is  because we wanted to ensure sufficient
time for system relaxation after $\lambda$-parameter
exchange. 
  
In REM1 there are 16 replicas with 16 different temperatures listed in 
Table I.
The temperatures are distributed exponentially, 
following the optimal distribution found in Ref.~\cite{SO}.
After every 10 fs of parallel MD simulations, eight pairs of 
replicas corresponding to neighboring temperatures
were simultaneously exchanged, and the pairing was alternated between
the two possible choices \cite{SO}.
   
For umbrella potentials, the O1 to H5 
hydrogen-bonding distance, or ``end-to-end distance,''
was chosen 
as the reaction coordinate $\xi$ and the harmonic restraining potentials
of $\xi$ in Eq.~(\ref{Eqn26}) were imposed.
The force constants, $k_{\ell}$, and 
the midpoint positions, $d_{\ell}$, are listed in Table I. 
    
In REUS1 and US1, 14 replicas were simulated with the
same set of umbrella potentials at $T=300$ K. 
The first parameter value, 0.0 (0.0), in Table I means that the 
restraining potential is null, i.e., $V_{\ell}=0$.
The remaining 13 sets of parameters are the same as
those adopted in Ref.~\cite{Boc1}.
Let us order the umbrella potentials, $V_{\ell}$ in Eq.~(\ref{Eqn25}),
in the increasing order of the midpoint value $d_{\ell}$,
i.e., the same order that appears in Table I.
We prepared replicas so that the potential energy for each
replica includes exactly one umbrella potential
(here, we have $M = L = 14$).
Namely, in Eq.~(\ref{Eqn25}) for $\bil = \bil_m$ we set
\begin{equation}
\lambda^{(\ell)}_m = \delta_{\ell,m}~,
\label{Eqn31}
\end{equation}
where $\delta_{k,l}$ is Kronecker's delta function, and
we have
\begin{equation}
E_{\bil_m} (q^{[i]}) = E_0 (q^{[i]}) + V_m (q^{[i]})~.
\label{Eqn32}
\end{equation}
The difference between REUS1 and US1 is whether replica exchange
is performed or not during the parallel MD simulations.
In REUS1 seven pairs of 
replicas corresponding to ``neighboring'' umbrella potentials,
$V_{m}$ and $V_{m+1}$,
were simultaneously exchanged after every 200 fs of parallel
MD simulations, and the pairing was alternated between
the two possible choices.  (Other pairings will have much smaller 
acceptance ratios of replica exchange.)
The acceptance criterion for replica exchange is given
by Eq.~(\ref{Eqn20}), where Eq.~(\ref{Eqn21}) now reads
(with the fixed inverse temperature $\beta = 1/300 k_B$)
\begin{equation}
\Delta = \beta 
\left(V_m\left(q^{[j]}\right) - 
      V_m\left(q^{[i]}\right) -
      V_{m+1}\left(q^{[j]}\right) + 
      V_{m+1}\left(q^{[i]}\right)\right)~,
\label{Eqn33}
\end{equation}
where replica $i$ and $j$ respectively have umbrella potentials
$V_m$ and $V_{m+1}$ before the exchange.

In REUS2 and US2, 16 replicas were simulated at four different
temperatures with four different restraining potentials
(there are $L=4$
umbrella potentials at $N_T=4$ temperatures, making the total
number of replicas $M=N_T \times L=16$; see Table I).
We can introduce the following labeling for the parameters
characterizing the replicas:
\begin{equation}
\begin{array}{rl}
\biL_m = (T_m,\bil_m) & \longrightarrow
\ \biL_{I,J} = (T_I,\bil_J)~. \cr
(m=1, \cdots, M) & \ \ \ \ \ \ \ \ \ \ (I=1, \cdots, N_T,~J=1, \cdots, L)
\end{array}
\label{Eqn34}
\end{equation}
The potential energy is given by Eq.~(\ref{Eqn32})
with the replacement: $m \rightarrow J$.
Let us again order the umbrella potentials, $V_J$,
and the temperatures, $T_I$, in
the same order that appear in Table I.
The difference between REUS2 and US2 is 
whether replica exchange
is performed or not during the MD simulations.
In REUS2 we performed the following replica-exchange processes alternately
after every 200 fs of parallel MD simulations:
\begin{enumerate}
\item Exchange pairs of replicas corresponding to neighboring temperatures,
$T_I$ and $T_{I+1}$ 
(i.e., exchange replicas $i$ and $j$ that
respectively correspond to parameters
$\biL_{I,J}$ and $\biL_{I+1,J}$).
(We refer to this process as $T$-exchange.)
\item Exchange pairs of replicas corresponding to 
``neighboring'' umbrella potentials,
$V_J$ and $V_{J+1}$ 
(i.e., exchange replicas $i$ and $j$ that
respectively correspond to parameters
$\biL_{I,J}$ and $\biL_{I,J+1}$).
(We refer to this process as $\lambda$-exchange.)
\end{enumerate}
In each of the above processes, two pairs of replicas were simultaneously
exchanged, and the pairing was further 
alternated between the two possibilities.
The acceptance criterion for these replica exchanges is given by 
Eq.~(\ref{Eqn20}), where Eq.~(\ref{Eqn21}) now reads
\begin{equation}
\Delta = \left(\beta_{I} - \beta_{I+1} \right)
\left(E_0 \left(q^{[j]}\right) 
    + V_J \left(q^{[j]}\right) 
    - E_0 \left(q^{[i]}\right)
    - V_J \left(q^{[i]}\right)\right)~, 
\label{Eqn35}
\end{equation}
for $T$-exchange, and
\begin{equation}
\Delta = \beta_I 
\left(V_J\left(q^{[j]}\right) - 
      V_J\left(q^{[i]}\right) -
      V_{J+1}\left(q^{[j]}\right) + 
      V_{J+1}\left(q^{[i]}\right)\right)~,
\label{Eqn36}
\end{equation}
for $\lambda$-exchange.
By this procedure, the random walk 
in the reaction coordinate space as well as in temperature
space can be realized.
Note that we carry out the velocity rescaling of Eq.~({\ref{eqn11})
in $T$-exchange.  In principle, we can also introduce a similar velocity
rescaling in $\lambda$-exchange to the two relevant 
atoms O1 and H5 
in order to adjust for the exchange of the
restraining potentials  
(because the restraining force acts only on O1 and H5). 
We also incorporated this rescaling but
did not see much improvement
in performance.  The results presented 
below are thus those from no velocity rescaling in $\lambda$-exchange.
 
We now give the details of the results obtained in the present work.
First of all, we examine whether the replica-exchange processes 
properly occurred in REM and REUS simulations.  One criterion for
the optimal performance is:
whether the acceptance ratio of replica exchange is
uniform and sufficiently large or not.
In Tables II--IV we list
the acceptance ratios of replica exchange
corresponding to the adjacent pairs of temperatures or the restraining 
potentials. In all cases the acceptance ratios are almost uniform and
large enough ($> 10$ \%); all simulations indeed performed properly.
In particular, the acceptance ratios for exchanging adjacent 
temperatures are 
significantly uniform in all cases, implying that the
exponential temperature distributions of Ref.~\cite{SO}
are again optimal.
However, the acceptance ratios for exchanging adjacent restraining 
potentials are not perfectly uniform, and there is some room
for fine tuning.
The acceptance ratio for exchanging restraining potentials depends
on the strength of the force constants, $k_{\ell}$, 
(see Eq.~(\ref{Eqn26})) and we weakened the value from 1.2 
kcal/mol$\cdot$\AA$^2$ in REUS1 to 0.3 kcal/mol$\cdot$\AA$^2$ 
in REUS2 in order to have sufficient replica
exchanges in REUS2.  This is because we have a much 
smaller number of restraining
potentials in REUS2 than in REUS1 (3 versus 13), and yet both simulations
have to cover the same range of reaction coordinate $\xi$, i.e., 
from 0 \AA \  to 13.8 \AA. 

In order to have sufficient replica exchanges between neighboring
temperatures and between neighboring restraining potentials,
the probability distributions corresponding to
neighboring parameters should have enough overlaps.
In Fig.~1(a) the canonical probability distributions
of the unbiased potential energy $E_0$ at
the four chosen temperatures are shown.  The results are 
for the parameters $\biL_{I,1}$ $(I=1, \cdots, 4)$, i.e., for
the case of no restraining potentials, and were
obtained from the REUS2 simulation.  
In Fig.~1(b) the probability distributions of the reaction coordinate
$\xi$ with the four chosen restraining potentials are shown.
The results are 
for the parameters $\biL_{2,J}$ $(J=1, \cdots, 4)$, i.e.,
for the temperature $T = 315$ K, and were also obtained 
from the REUS2 simulation.  
In both figures we do observe sufficient overlaps in pairs of
the distributions corresponding to the
neighboring parameter values, and this is reflected in
the reasonable acceptance ratios listed in Table IV. 

In order to further confirm that our REM simulations performed properly,
we have to examine the time series of various quantities and
observe random walks.  For instance,
in Fig.~2 the trajectories of a few quantities in REUS2 are shown. 
In Fig.~2(a) we show the time series of replica exchange for
the parameter $\biL_{1,1} = (T_1,\bil_1)$ (i.e.,
$T_1=250$ K and $k_1 = d_1 = 0.0$).
We do observe a random walk in replica space, and we see that
all the replicas frequently visited the parameter value $\biL_{1,1}$.
  
The complementary picture to this is the time series of
$T$-exchange and $\lambda$-exchange for each replica.
Free random walks  
both in ``temperature space'' and in ``restraining 
potential space''
were indeed observed. 
For instance, the time series of temperature exchange 
for one of the replicas
(replica 1) is shown in Fig.~2(b).
The corresponding time series of the reaction coordinate $\xi$,
the distance between atoms O1 and H5, for the same replica is shown 
in Fig.~2(c).  We see that the conformational
sampling along the reaction coordinate is significantly enhanced. 
In the blocked alanine trimer, the reaction coordinate $\xi$ can be
classified into three regions \cite{Boc1}: 
the helical region ($\xi < 3$ \AA), the turn region
(3 \AA\ $< \xi <$ 7 \AA), and the extended region
($\xi >$ 7 \AA).
Thus, Fig.~2(c) implies that helix-coil transitions
frequently occurred during the replica-exchange simulation, whereas
in the conventional canonical 
simulations such a frequent folding and unfolding process cannot be seen.
 
After confirming that the present REM and REUS simulations
performed properly, we now present and compare the
physical quantities
calculated by these simulations.
In Fig.~3 the potentials of mean force (PMF) of the
unbiased system along the 
reaction coordinate $\xi$ at $T=300$ K are shown.
The results are from REM1, REUS1, and US1 simulations.
For these calculations, the WHAM equations of Eqs.~(\ref{Eqn27})
and (\ref{Eqn28}) were solved by iteration
first, and then Eq.~(\ref{Eqn30}) was used to obtain the PMF.
We remark that for biomolecular systems the results obtained from the
WHAM equations are insensitive to the values of $g_m$ in
Eq.~(\ref{Eqn27}) \cite{WHAM}.  Hence, we set 
$g_m=$ constant in the present article.
From Fig.~3 we see that the 
PMF curves obtained by REM1 and REUS1 are 
essentially identical for low values of $\xi$ ($\xi < 7$ \AA).
The two PMF curves start deviating slightly,
as $\xi$ gets larger, and for $\xi > 9$ \AA \ 
the agreement completely deteriorates.
The disagreement comes from the facts that
the average $\xi$ at the highest temperature in REM1 
($T_{16}=1500$ K) 
is $<\xi>_{T_{16}} = 8.0$ \AA \ and that the original REM
with $T$-exchange only cannot sample accurately the region
where $\xi$ is much larger than 
$<\xi>_{T_{16}}$.
These two simulations were performed under very different
conditions: One was run at different temperatures without
restraining potentials and the other at one temperature
with many restraining potentials (see Table I).
We thus consider the results to be quite reliable for
($\xi < 9$ \AA).
   
On the other hand, the PMF obtained by US1 is relatively 
larger than those obtained by REM1 and REUS1 in the region 
of 2 \AA\ $< \xi <$ 4 \AA, which corresponds to 
the structural transition state between the $\alpha$-helical 
and turn structures.  This suggests that
US1 got trapped in states of energy local
minima at $T=300$ K. In the region of completely extended structures
($\xi > 9$ \AA), the results of REUS1 and US1 are similar
but the discrepancy is again non-negligible.
We remark that at $T=300$ K the
PMF is the lowest for $\xi = 2$ \AA, which
implies that the $\alpha$-helical structure is favored at this 
temperature.
 
We next study
the temperature dependence of physical quantities obtained from the
REM1, REUS2, and US2 simulations.
In Fig.~4(a) we show the PMF again at $T=300$ K.  We observe that
the PMF curves from REM1 and REUS2 are essentially identical
for $\xi < 9$ \AA \ and that they deviate
for $\xi > 9$ \AA, because the results for REM1 are not reliable
in this region as noted above.
In fact, by comparing Figs.~3 and 4(a), we find that the PMF
obtained from REUS1 and REUS2 are almost in complete agreement
at $T=300$ K in the entire
range of $\xi$ values shown.  On the other hand, we
observe a discrepancy between REUS2 and US2 results.
The PMF curve for US2 is significantly less than that for
REUS2 in the region 2 \AA\ $<\xi<$ 8 \AA.  Note that the 
PMF curves for US1 and US2 are completely in disagreement
(compare Figs.~3 and 4(a)).

In Fig.~4(b) we show the PMF at $T=500$ K, which
we obtained from REM1, REUS2, and US2 simulations.
We again observe that the results from REM1 and REUS2
are in good agreement for a wide range of $\xi$ values.
We find that the results from REM1 do not significantly deteriorate
until $\xi>11$ \AA \ at $T=500$ K, whereas it did start
deviating badly for $\xi >9$ \AA \ at $T=300$ K.  The PMF
curve for US2 deviates strongly from the REUS2 results
for $\xi>6$ \AA \ and is much larger than that of REUS2
(and REM1) in this region.
We remark that at $T=500$ K the
PMF is the lowest for $\xi \approx 6$ \AA,
which implies that extended structures are favored at
this temperature.

In Fig.~5 we show the average values of the reaction
coordinate $\xi$ as a function of temperature.  The results
are again from the REM1, REUS2, and US2 simulations.
The expectation values were calculated from
Eq.~(\ref{Eqn29}).
We find that the average reaction coordinate, or the
average end-to-end distance, grows as the temperature is raised,
reflecting the unfolding of the peptide upon increased
thermal fluctuations.
Again we observe an agreement between REM1 and REUS2,
whereas the results of US2 deviate.

Let us emphasize that the total length of the MD simulations
was the same (2 ns) for each replica in all the simulations
performed.  Hence, we have shown that the replica-exchange
umbrella sampling can give much more accurate free-energy
profiles along a reaction coordinate than the conventional umbrella sampling.
 
\section{CONCLUSIONS}
In this article we have presented a multidimensional extension of
the original replica-exchange method.  One example of this approach
is the combination of the replica-exchange method with the umbrella
sampling, which we refer to as 
the {\it replica-exchange umbrella sampling} (REUS).
While pairs of replicas with different
temperatures are exchanged during the simulation in the original
replica-exchange method, pairs of replicas with different temperatures
and/or different biasing potentials for the umbrella sampling are
exchanged in REUS.
This greatly enhances the sampling of the conformational
space and allows accurate calculations of free energy in a wide
temperature range from a single simulation run, using the
weighted histogram analysis method.
The difference between REUS and the conventional umbrella sampling
is just whether the replica-exchange process is performed or not. 
Only minor modifications to
the conventional umbrella sampling method are necessary. 
However, the advantage of REUS over the umbrella sampling is 
significant, and the effectiveness was established with the
system of an alanine trimer.

%

\vspace{0.5cm}
\noindent
{\bf Acknowledgments:} \\
Our simulations were performed on the Hitachi and other computers at the
IMS Computer Center. A part of the computation was done also at the Center for
Promotion of Computational Science and Engineering of Japan Atomic Energy
Research Institute and Computer Center of the National Institute of Genetics.
This work is supported, in part, by a grant from the Research for the Future 
Program of the Japan Society for the Promotion of Science 
(Grant No. JSPS-RFTF98P01101).

%
%

\begin{table}
\caption{Summary of the replica parameters for the present simulations.}
\begin{center}
\begin{tabular}{lcrlrl}
Run$^a$ & $M^b$ & ${N_T}^b$ & ~~Temperature [K] & $L^b$ & 
~~~~~~~$d_{\ell}$ [\AA] $(k_{\ell}$ [kcal/mol$\cdot$\AA$^2])^c$ \\
  \hline
REM1 & 16 & 16 & 200, 229, 262, 299, & 0 &  \\	
       &    &    & 342, 391, 448, 512, &   &   \\
       &    &    & 586, 670, 766, 876, &   &   \\
       &    &    & 1002, 1147, 1311, &   &   \\
       &    &    & 1500 &   &   \\
REUS1, US1 & 14 & 1 & 300 & 14 & 0.0 (0.0)$^d$, 1.8 (1.2), 2.8 (1.2), 3.8 (1.2),\\
       &    &   &     &    & 4.8 (1.2), 5.8 (1.2), 6.8 (1.2), 7.8 (1.2),\\
       &    &   &     &    & 8.8 (1.2), 9.8 (1.2), 10.8 (1.2), 11.8 (1.2),\\
       &    &   &     &    & 12.8 (1.2), 13.8 (1.2)\\  
REUS2, US2 & 16 & 4 & 250, 315, 397, 500 & 4 & 0.0 (0.0), 7.8 (0.3), 10.8 (0.3), 13.8 (0.3)\\
\end{tabular}
\end{center}
\noindent
$^a$ REM, REUS, and US stand for an original replica-exchange
simulation, replica-exchange umbrella sampling simulation, and
conventional umbrella sampling simulation, respectively.
  
\noindent
$^b$ $M$, $N_T$, and $L$ are the total numbers of replicas,
temperatures, and restraining potentials, respectively 
(see Eqs.~(\ref{Eqn19}) and (\ref{Eqn25})).  In REUS2 and
US2 we set
$M=N_T \times L$ for simplicity.  We remark that this
relation is not always required.  For instance,
the 16 replicas could have 16 different temperatures with 16
different restraining potentials (i.e., $M=N_T=L=16$).

\noindent
$^c$ $d_{\ell}$ and $k_{\ell}$ $(\ell = 1, \cdots, L)$ 
are the strengths and the midpoints
of the restraining potentials, respectively (see Eq.~(\ref{Eqn26})).

\noindent
$^d$ The parameter value 0.0 (0.0) means that the restraining
potential is null, i.e., $V_{\ell}=0$.
\label{Tab1}
\end{table}%


\begin{table}
\caption{Acceptance ratios of replica exchange in REM1.}
\begin{center}
\begin{tabular}{cc}
  Pair of Temperatures & Acceptance Ratio \\
  \hline
  200 $\longleftrightarrow$ 229 &  0.430 \\
  229 $\longleftrightarrow$ 262 &  0.433 \\
  262 $\longleftrightarrow$ 299 &  0.433 \\
  299 $\longleftrightarrow$ 342 &  0.428 \\
  342 $\longleftrightarrow$ 391 &  0.430 \\
  391 $\longleftrightarrow$ 448 &  0.423 \\
  448 $\longleftrightarrow$ 512 &  0.429 \\
  512 $\longleftrightarrow$ 586 &  0.427 \\
  586 $\longleftrightarrow$ 670 &  0.434 \\
  670 $\longleftrightarrow$ 766 &  0.437 \\
  766 $\longleftrightarrow$ 876 &  0.445 \\
  876 $\longleftrightarrow$ 1002 &  0.446 \\
  1002 $\longleftrightarrow$ 1147 &  0.446 \\
  1147 $\longleftrightarrow$ 1311 &  0.454 \\
  1311 $\longleftrightarrow$ 1500 &  0.452 \\
\end{tabular}
\end{center}
\label{Tab2}
\end{table}%


\begin{table}
\caption{Acceptance ratios of replica exchange in REUS1.}
\begin{center}
\begin{tabular}{cc}
  Pair of Restraint Parameters & Acceptance Ratio \\
  \hline
  0.0 (0.0) $\longleftrightarrow$ 1.8 (1.2) &  0.202 \\
  1.8 (1.2) $\longleftrightarrow$ 2.8 (1.2) &  0.210 \\
  2.8 (1.2) $\longleftrightarrow$ 3.8 (1.2) &  0.174 \\
  3.8 (1.2) $\longleftrightarrow$ 4.8 (1.2) &  0.161 \\
  4.8 (1.2) $\longleftrightarrow$ 5.8 (1.2) &  0.223 \\
  5.8 (1.2) $\longleftrightarrow$ 6.8 (1.2) &  0.155 \\
  6.8 (1.2) $\longleftrightarrow$ 7.8 (1.2) &  0.211 \\
  7.8 (1.2) $\longleftrightarrow$ 8.8 (1.2) &  0.229 \\
  8.8 (1.2) $\longleftrightarrow$ 9.8 (1.2) &  0.119 \\
  9.8 (1.2) $\longleftrightarrow$ 10.8 (1.2) &  0.276 \\
  10.8 (1.2) $\longleftrightarrow$ 11.8 (1.2) &  0.156 \\
  11.8 (1.2) $\longleftrightarrow$ 12.8 (1.2) &  0.138 \\
  12.8 (1.2) $\longleftrightarrow$ 13.8 (1.2) &  0.383 \\
\end{tabular}
\end{center}
\label{Tab3}
\end{table}%


\begin{table}
\caption{Acceptance ratios of replica exchange in REUS2.}
\begin{center}
\begin{tabular}{ccc}
  Temperature & Pair of Restraint Parameters & Acceptance Ratio \\
  \hline
  250 & 0.0 (0.0)  $\longleftrightarrow$ 7.8 (0.3)  &  0.149 \\
  250 & 7.8 (0.3)  $\longleftrightarrow$ 10.8 (0.3) &  0.104 \\
  250 & 10.8 (0.3) $\longleftrightarrow$ 13.8 (0.3) &  0.127 \\
  315 & 0.0 (0.0)  $\longleftrightarrow$ 7.8 (0.3)  &  0.250 \\
  315 & 7.8 (0.3)  $\longleftrightarrow$ 10.8 (0.3) &  0.105 \\
  315 & 10.8 (0.3) $\longleftrightarrow$ 13.8 (0.3) &  0.120 \\
  397 & 0.0 (0.0)  $\longleftrightarrow$ 7.8 (0.3)  &  0.363 \\
  397 & 7.8 (0.3)  $\longleftrightarrow$ 10.8 (0.3) &  0.135 \\
  397 & 10.8 (0.3) $\longleftrightarrow$ 13.8 (0.3) &  0.160 \\
  500 & 0.0 (0.0)  $\longleftrightarrow$ 7.8 (0.3)  &  0.483 \\
  500 & 7.8 (0.3)  $\longleftrightarrow$ 10.8 (0.3) &  0.185 \\
  500 & 10.8 (0.3) $\longleftrightarrow$ 13.8 (0.3) &  0.228 \\
  \hline
  Restraint Parameters & Pair of Temperatures & Acceptance Ratio \\
  \hline
  0.0 (0.0) & 250 $\longleftrightarrow$ 315 & 0.193 \\
  0.0 (0.0) & 315 $\longleftrightarrow$ 397 & 0.186 \\
  0.0 (0.0) & 397 $\longleftrightarrow$ 500 & 0.189 \\
  7.8 (0.3) & 250 $\longleftrightarrow$ 315 & 0.174 \\
  7.8 (0.3) & 315 $\longleftrightarrow$ 397 & 0.179 \\
  7.8 (0.3) & 397 $\longleftrightarrow$ 500 & 0.190 \\  
  10.8 (0.3) & 250 $\longleftrightarrow$ 315 & 0.189 \\
  10.8 (0.3) & 315 $\longleftrightarrow$ 397 & 0.184 \\
  10.8 (0.3) & 397 $\longleftrightarrow$ 500 & 0.195 \\  
  13.8 (0.3) & 250 $\longleftrightarrow$ 315 & 0.185 \\
  13.8 (0.3) & 315 $\longleftrightarrow$ 397 & 0.184 \\
  13.8 (0.3) & 397 $\longleftrightarrow$ 500 & 0.205 \\  
\end{tabular}
\end{center}
\label{Tab4}
\end{table}%

%
%

\newpage
\centerline{\bf Figure Captions}

\begin{itemize}
\item Fig.~1. Probability distributions obtained from
the replica-exchange umbrella sampling simulation (REUS2). 
(a) The canonical probability distributions
of the unbiased potential energy $E_0$ at
the four chosen temperatures (the curves from left to right
correspond to
$T=$ 250, 315, 397, 500 K).  The results are
for the parameters $\biL_{I,1}$ $(I=1, \cdots, 4)$, i.e., for
the case of no restraining potentials (see Table I).
(b) The probability distributions of the reaction coordinate
$\xi$, the distance between the atoms O1 and H5, with the four 
chosen restraining potentials
(the curves from left to right correspond to 
$d_{\ell}$ [\AA] $(k_{\ell}$ [kcal/mol$\cdot$\AA$^2]) = $
0.0 (0.0), 7.8 (0.3), 10.8 (0.3), 13.8 (0.3)).
The results are
for the parameters $\biL_{2,J}$ $(J=1, \cdots, 4)$, i.e.,
for the temperature $T = 315$ K (see Table I).
\item Fig.~2. Time series of the replica-exchange umbrella sampling
simulation (REUS2). 
(a) Replica exchange for
the parameter $\biL_{1,1} = (T_1,\bil_1)$ (i.e.,
$T_1=250$ K and $k_1 = d_1 = 0.0$, see Table I).
(b) Temperature exchange for one of the replicas (replica 1).
(c) The reaction coordinate $\xi$ for one of the replicas (replica 1).
\item Fig.~3. The PMF along the reaction coordinate $\xi$
at $T=300$ K.
The dotted, solid, and dashed curves were obtained from the
original REM (REM1),
the replica-exchange umbrella sampilng (REUS1), and the conventional
umbrella sampling (US1), respectively.
\item Fig.~4. The PMF along the reaction coordinate $\xi$
at two temperatures.
(a) The PMF at $T=300$ K. The dotted, solid, and dashed curves were obtained 
from the original REM (REM1), the replica-exchange umbrella sampling
(REUS2), and the conventional umbrella sampling (US2), respectively.
(b) The PMF at $T=500$ K. The dotted, solid, and dashed curves were obtained 
from the original REM (REM1), the replica-exchange umbrella sampling
(REUS2), and the conventional umbrella sampling (US2), respectively.
\item Fig.~5. Average values of the reaction coordinate $\xi$
as a function of temperature.
The dotted, solid, and dashed curves were obtained from
the original REM (REM1),
the replica-exchange umbrella sampling (REUS2), and the conventional
umbrella sampling (US2), respectively.
Although the highest temperature in REM1 is 1500 K, only the results
for the temperature range between 200 K and 1000 K are shown for
REM1.
Since the lowest and highest tempeatures in REUS2 and US2
are respectively 250 K and 500 K,
only the results between these temperatures are shown for
these simulations.
\end{itemize}

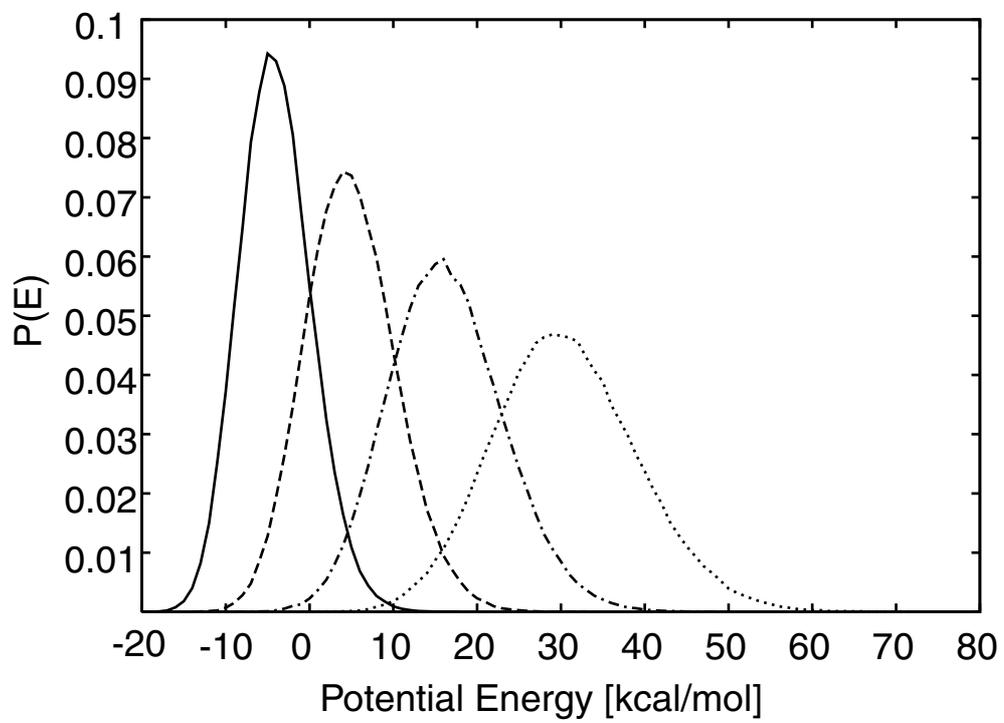
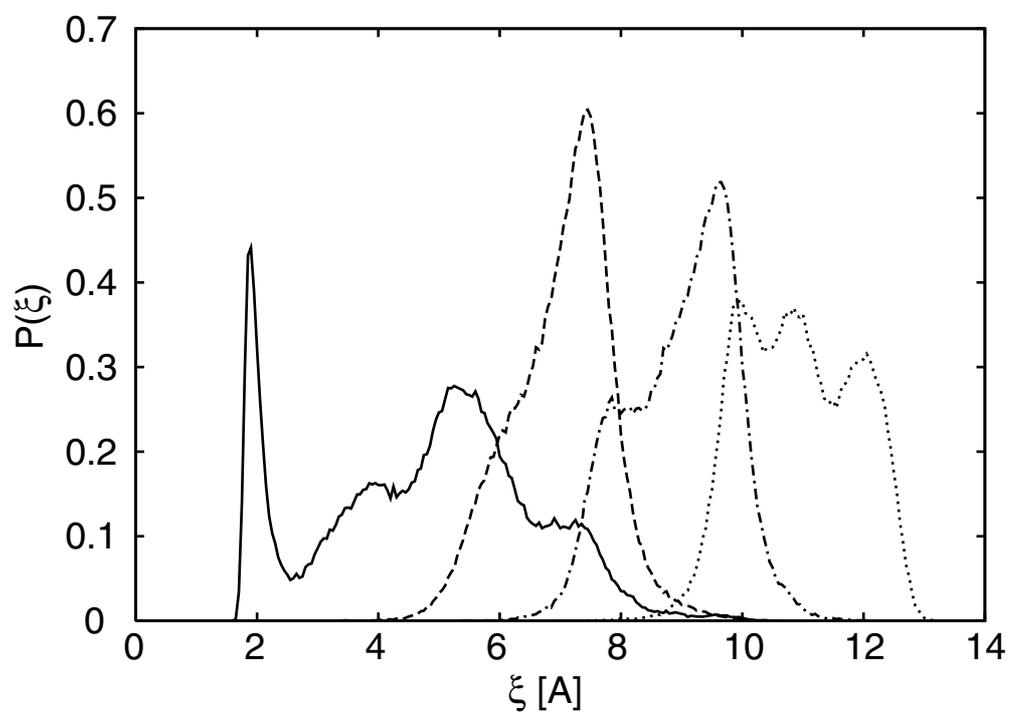
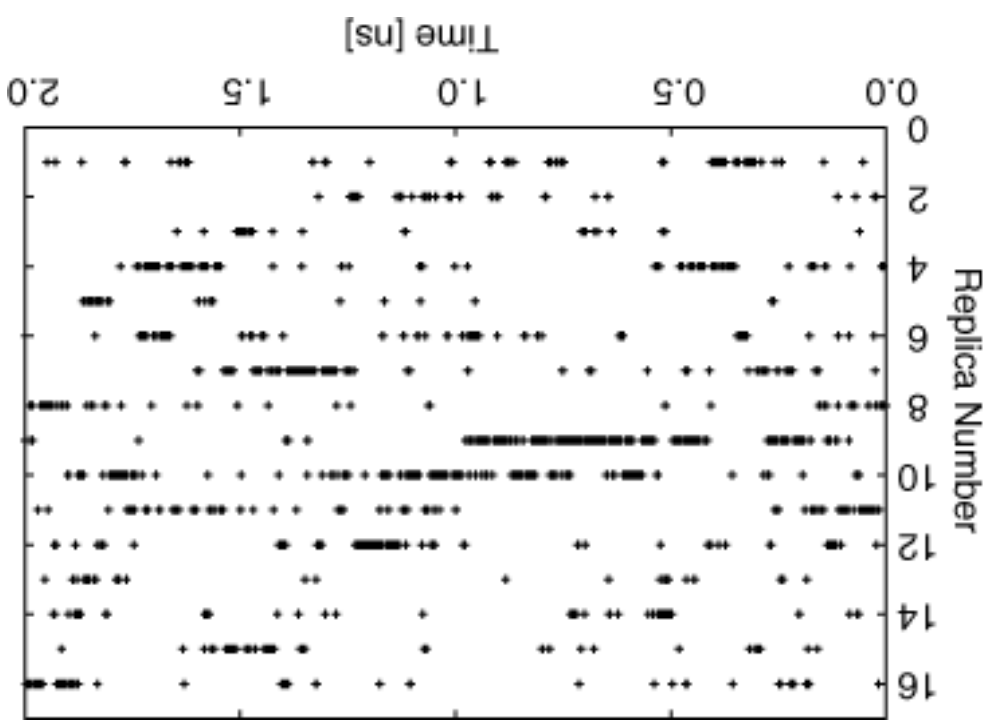
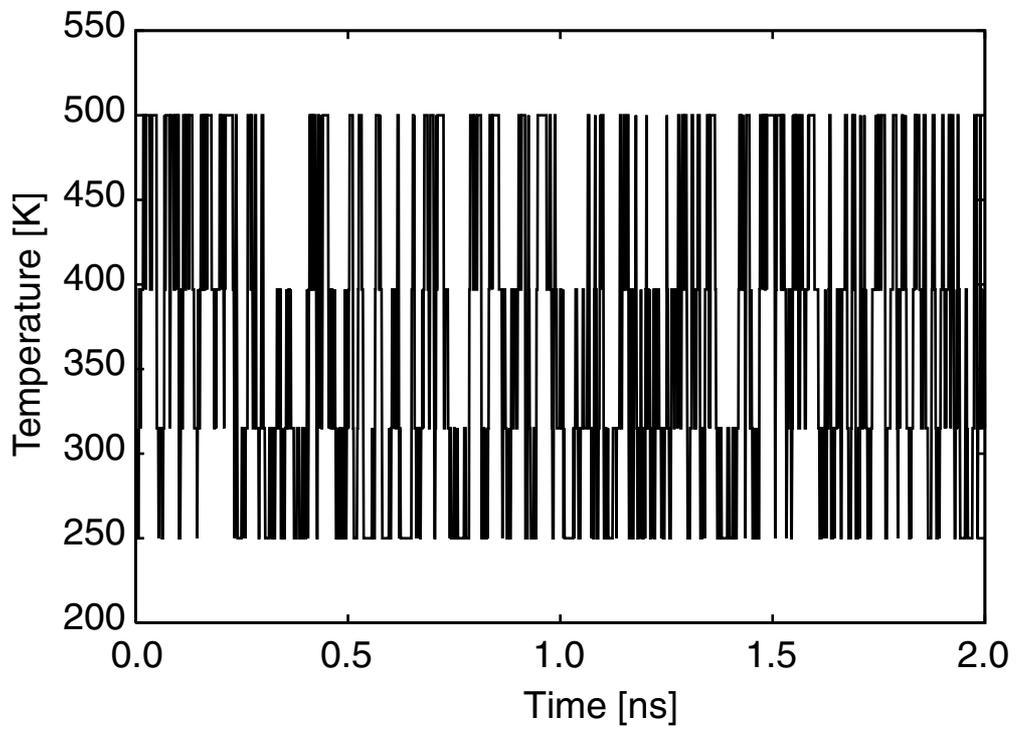
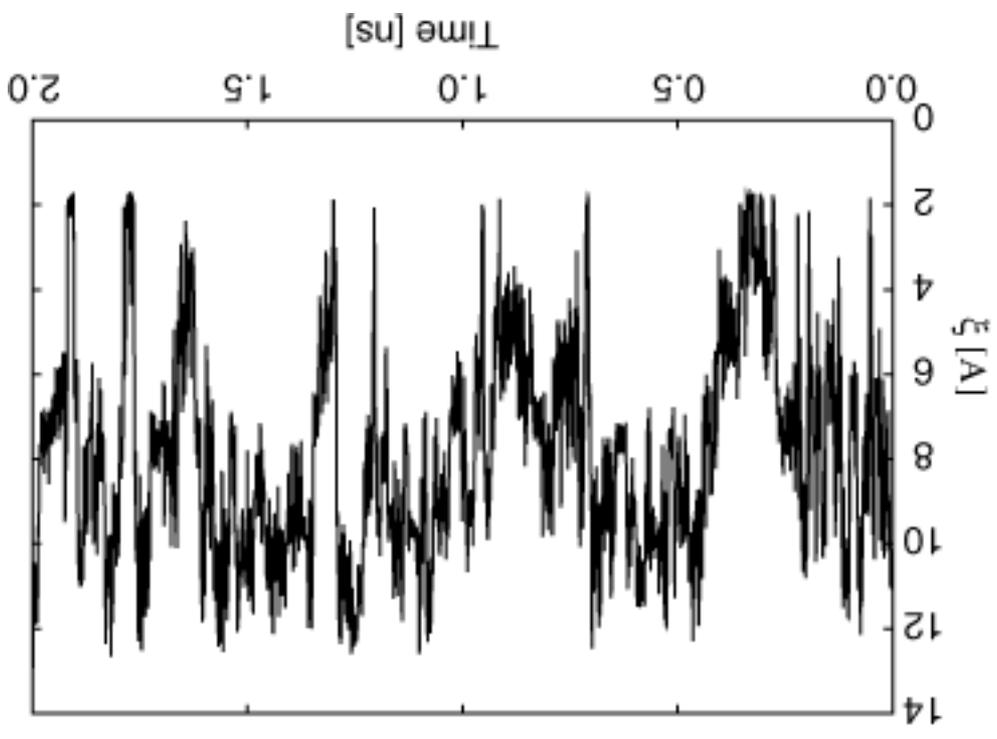
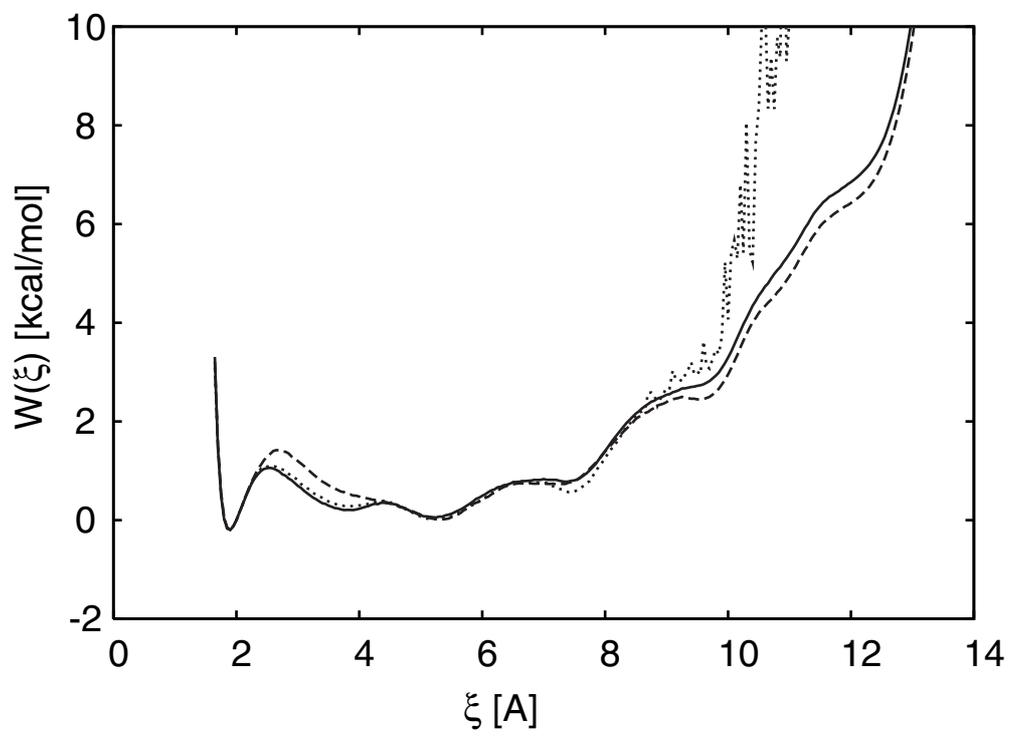
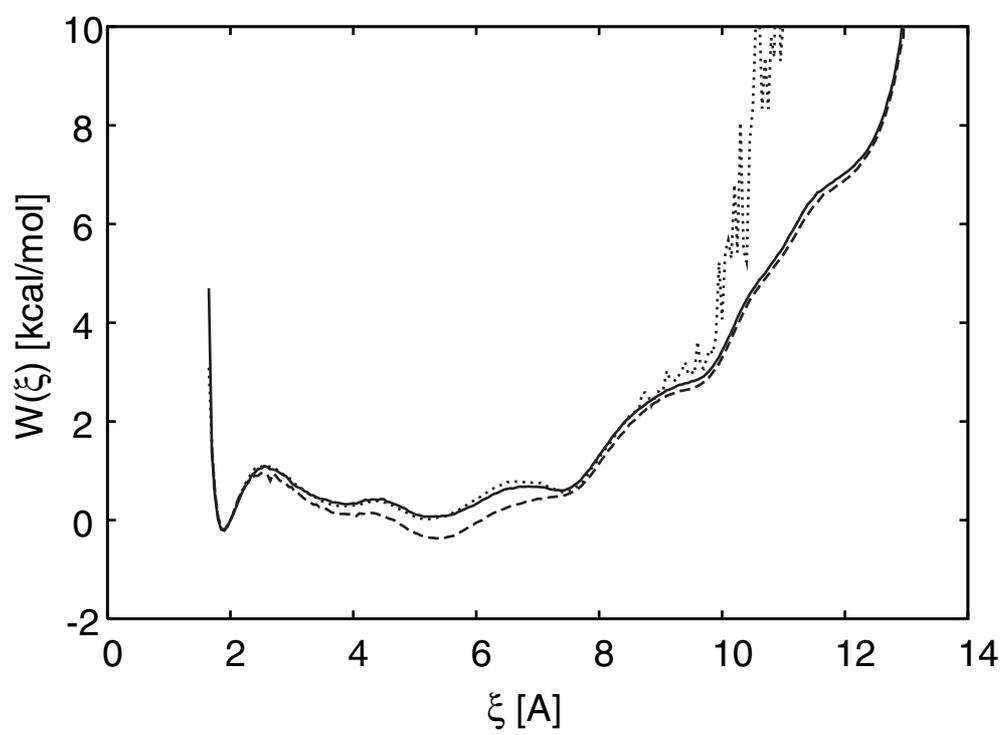
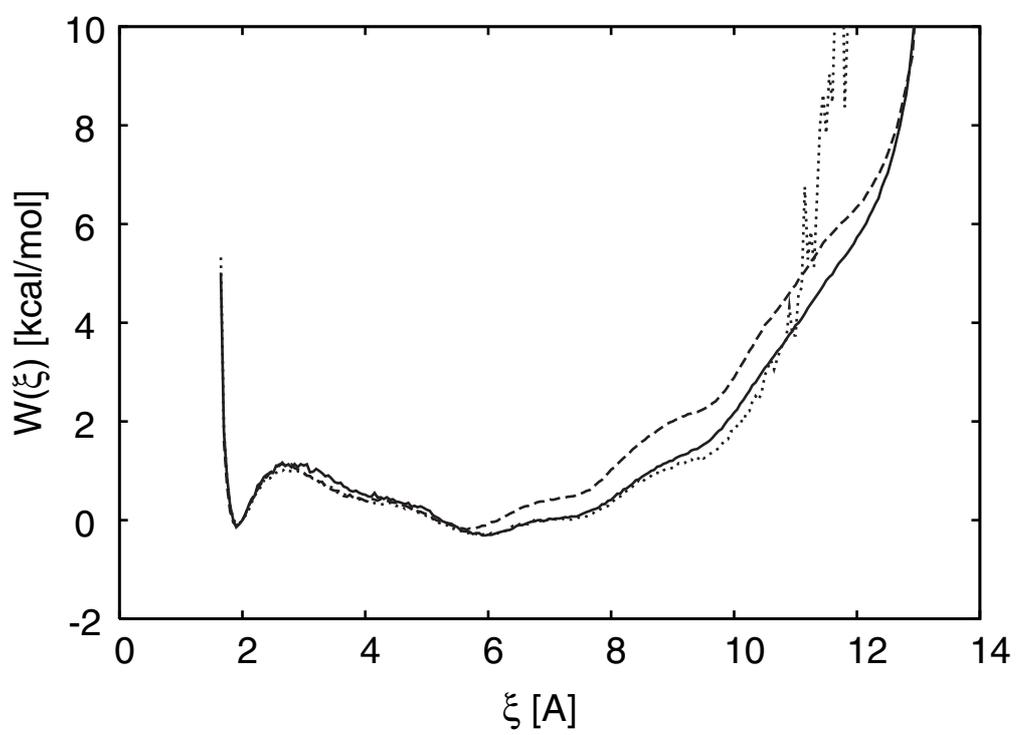
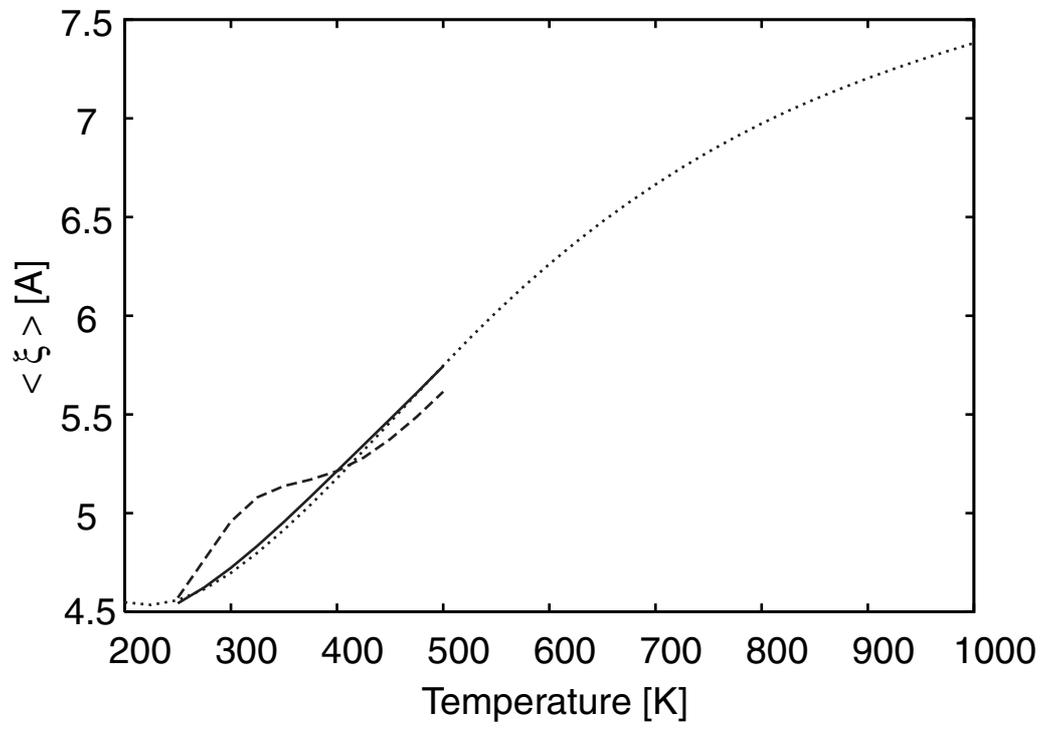
\end{document}